\documentclass[prl,preprint,showpacs]{revtex4}
\usepackage{amsmath,amssymb}
\usepackage{graphicx}
\usepackage{units}
\usepackage{latexsym}
\usepackage{bm}
\usepackage{color}

\begin{document}

\title{Observation of Electron Energy Discretization in Strong Field Double Ionization}

\author{
K. Henrichs$^1$, M. Waitz$^1$, F. Trinter$^1$, H. Kim$^1$, A. Menssen$^1$, H. Gassert$^1$, H. Sann$^1$,  T. Jahnke$^1$, J. Wu$^2$, M. Pitzer$^1$, M.
Richter$^1$, M. S. Sch\"offler$^1$, M. Kunitski$^1$
and R. D\"orner$^1$ }

\affiliation{ $^1$ Institut f\"ur Kernphysik,
J.~W.~Goethe-Universit\"at, Max-von-Laue-Str.1,
60438 Frankfurt am Main, Germany\\
$^2$State Key Laboratory of Precision
Spectroscopy, East China Normal University,
Shanghai 200062, China}


\pacs{32.80.Rm, 32.80.Fb, 32.90.+a, 42.50.Hz}
\date{\today}
\begin{abstract}
We report on the observation of discrete structures in the electron energy distribution for strong field \textit{double} ionization of Argon at 394 nm. The experimental conditions were chosen in order to ensure a non-sequential ejection of both electrons with an intermediate rescattering step. We have found discrete ATI (above-threshold ionization) like peaks in the sum energy of both electrons, as predicted by all quantum mechanical calculations. More surprisingly however is the observation of two ATI combs in the energy distribution of the individual electrons.

\end{abstract}

\maketitle

\textit{Single} ionization of atoms and molecules by
a multi-cycle short laser pulse leads to a
sequence of equidistant discrete peaks in the
energy spectrum of the ejected electron. These
above threshold ionization (ATI) peaks result
from the discretization of the photon field and
show that more photons than necessary to overcome
the binding can be absorbed from the field \cite{Agostini79prl}.
Similar ATI structure is predicted for non-sequential
\textit{double} ionization. For double ionization it is however the sum
energy of both electrons which is expected to
show the discretization \cite{Lein01pra,Liao10pra,Wang12pra,parker06prl,Armstrong11njp,Parker01jpb}.

At certain laser intenisities strong field double ionization is orders of
magnitude more efficient than expected from a
sequential scenario, where the atom is ionized by
subsequent independent interactions with the laser
field \cite{Walker94prl}. There is overwhelming
experimental and theoretical evidence that the
mechanism for this efficient double ejection
involves an intermediate step of recollision (see
\cite{Doerner02advances,Becker12rmp} for
reviews). This scenario is therefore termed non-sequential double ionization.
Initially only one electron is released. It
is accelerated by the laser field, driven back and
shares its energy with another bound electron.
The details of what exactly happens upon
recollision are under discussion. In addition to
evidence for a direct knock-out of a second
electron in an (e,2e) type collision
\cite{Doerner02advances}, there are also
contributions to the double ionization yield
where the second yet bound electron is excited upon
recollision with the first one and field ionized later during the laser
pulse (RESI) \cite{Feuerstein01prl}. At lower laser intensities there is
the suggestion that an intermediate
doubly excited transition state is formed upon
recollision \cite{sacha_pathways_2001,Camus12prl}. One possible pathway to create such a transition state is subsequent multiple, inelastic field-assisted recollisions, which resonantly excite the target ion \cite{Ho2005prl,Liu2008prl,Liu2010prl}.

Models based on classical physics have proven to be
highly successful in reproducing most of the
observed features \cite{Becker12rmp}. They do
however neglect the discrete photon nature of the
radiation field. Hence all energies of the
electrons are continuous in these classical
calculations. All quantum models of strong field ionization
ionization, on the contrary, predict a
discretization of energy in the final state
continuum (see e.g.
\cite{Lein01pra,Liao10pra,Wang12pra,parker06prl,Armstrong11njp,Parker01jpb}).
In a time dependent picture energy discretization
arises from the periodicity of the ionization
events in time
\cite{Lindner05prl,Armstrong11njp}. This concept
has been generalized to the two electron case
\cite{Armstrong11njp} where the sum energy of
both electrons shows discrete peaks. The ATI
structure on the electron sum energy was
predicted already in the first one-dimensional quantum
simulations of strong field double ionization
\cite{Lein01pra,Parker01jpb}. It naturally occurs
in all quantum models when more than two cycles of
the field are taken into account. The energy of
these ATI structures in double ionization is
predicted to be \cite{Lein01pra,Armstrong11njp}:

\begin{eqnarray}
E_1 + E_2 = n h\nu - I_{p1} - I_{p2} - 2U_p
\label{eqndouble}
\end{eqnarray}

Here $E_1, E_2$ are the continuum energies of the
two electrons, $n$ is the number of absorbed
photons with an energy of $h \nu$. $I_{p1}$ and $I_{p2}$
are the Stark shifted ionization potentials of
the neutral and the singly charged atom, respectively (for Argon
$I_{p1}=\unit[15.76]{eV}$ and $I_{p2}=\unit[27.63]{eV}$). The
ground state Stark shift is usually small. $U_p$
is the ponderomotive potential which is the AC
stark shift of the continuum.

On the experimental side, however, the predicted discretization of the electron sum
energy in non-sequential double ionization has evaded observation so far.
One possible reason for this is the focal
averaging in the experiment. As $U_p$ changes
with intensity across the focus the focal
averaging smears out the discrete peaks (see eq. 1). To
circumvent this problem we have chosen \unit[394]{nm}
light for our present study. This doubles the
spacing between ATI peaks compared to the fundamental Ti:Sa output and at the same time
reduces $U_p$. The \unit[394]{nm} pulses were generated
in a $\unit[200]{\mu m}$ thick BBO crystal by frequency
doubling of the \unit[788]{nm} output of a Ti:Sa laser system (\unit[100]{kHz},
$\unit[100]{\mu J}$, \unit[45]{fsec}, Wyvern-500, KMLabs).
A COLTRIMS reaction microscope \cite{Ullrich03rpp} was used
to measure both electrons in coincidence
with the doubly charged Ar ion.
The gas density in our jet was adjusted
to reach a count rate of
\unit[13]{kHz} on the electron detector and an ion count rate of \unit[2.5]{kHz}, 70\% of which was Argon. We used momentum conservation between the
two detected electrons and the ion to suppress events coming from
false coincidences. This aproach however does not eliminate false coincidences completely due to the finite momentum resolution. One kind of false coincidences that remain is when one of the two detected electrons originates from
ionization of a second atom in the same pulse. According to our estimations these events account for
less than 20 \%. We have
generated a false correlated electron energy
spectrum of these false coincidences and subsequently subtract it from the
raw data. The laser intensity in the interaction region has been determined by measuring
the shift of the sum energy of electron and
proton from dissociative ionization of $H_2$.
This shift, which is given by $U_p$, has been found
to be linear with laser power. We estimate the
accuracy of our calibration to be better than $\pm$20 \%.

\begin {figure}[t]
  \begin{center}
    \includegraphics[width=0.5\linewidth]{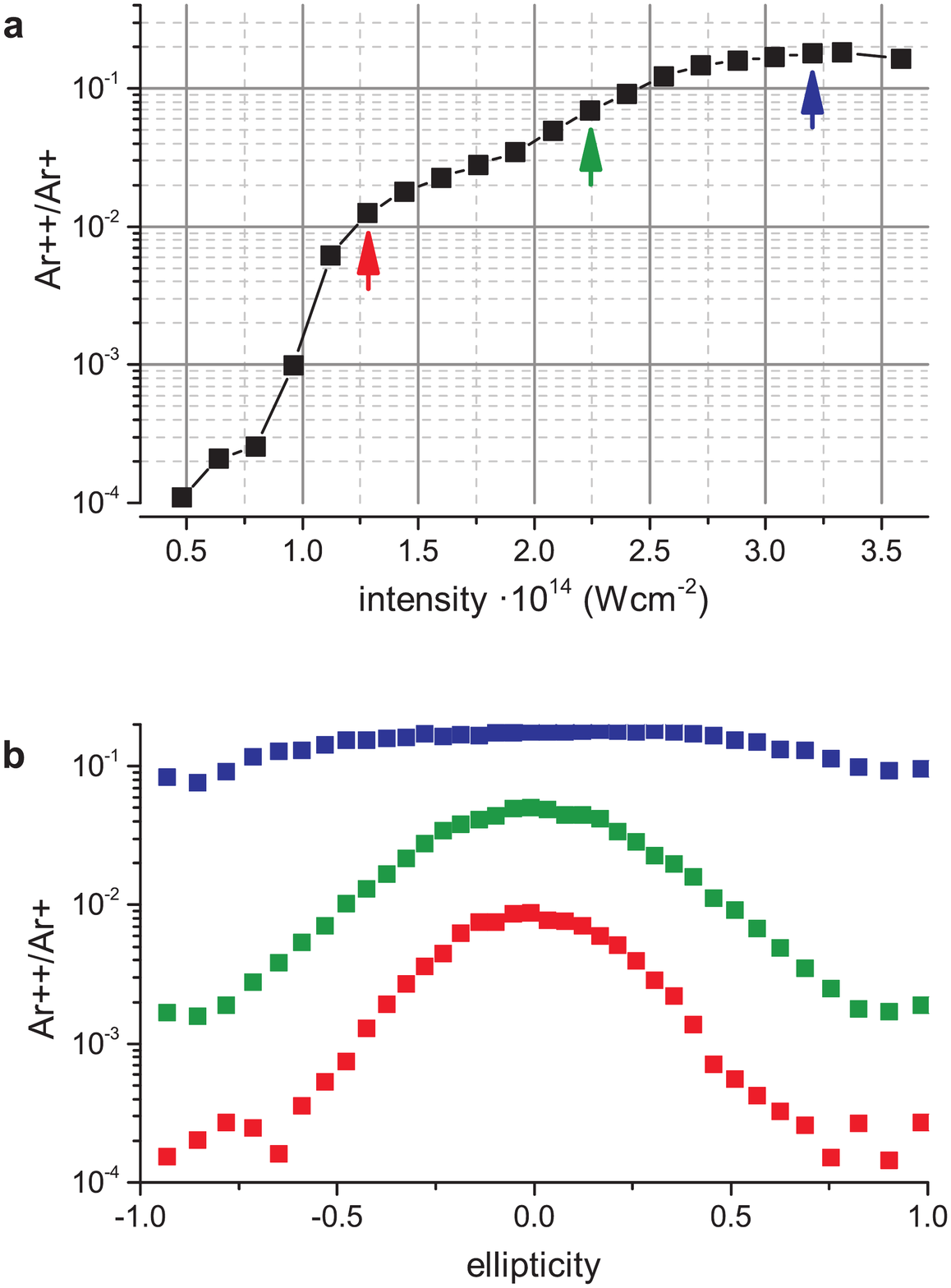}
    \caption{Ratio of double to single ionization of Ar by \unit[394]{nm}, \unit[45]{fsec} laser pulses.
    (a) as function of laser intensity  (b) as function of polarization ellipticity at the intensity marked by color arrows in (a). The intensity shown by the red arrow was used in the experiment ($\unit[1.3\times10^{14}]{Wcm^{-2}}$). Note the logarithmic scale.}
  \label{fig1}
  \end{center}
\end {figure}

We first confirm recollision to be responsible for double ionization at the parameters of our experiment.
Figure 1a shows the intensity dependence of the
double ionization probability. It rises only slowly
over a range of intensities ($\unit[1.2-2.2\times10^{14}]{Wcm^{-2}}$, referred to as "the
knee" in the literature) because the ejection of
the second electron is driven by the
electron-electron interaction and not directly by
the field. The ratio drops steeply at lower
intensities, when the energy of the recolliding
electron is insufficient to contribute
significantly to ejection of the second electron.
These general features of the double ionization
probability are very similar to those known for
\unit[800]{nm}. For the experiment we choose an intensity of $\unit[1.3\times10^{14}]{Wcm^{-2}}$ (shown with a red arrow in Figure 1a), just at the onset of the non-sequential regime. Under this condition the Keldysh parameter \cite{Keldysh1965} is 2, indicating a multi-photon regime of ionization. The chosen laser intensity corresponds to $U_p\approx\unit[1.9]{eV}$, i.e. the maximum energy of the recolliding electron \cite{Corkum93prl} is
$3.17\times{U_p}\approx\unit[6]{eV}$. This energy is much below the second ionization
potential $I_{p2}=\unit[27.6]{eV}$ and does not even
cover the excitation energy of the first excited
state $(3s3p^6)$, which is \unit[13.5]{eV} \cite{Nist}. Thus, for this intensity the (e,2e) knock-out ionization and RESI should be excluded.
One of the possible double ionization scenarios in this case is recapture of the first electron under recollision. The released energy might now be enough for excitation of the second electron, which results in formation of a doubly excited Coulomb complex or compound state \cite{sacha_pathways_2001,Camus12prl}. It was shown that under certain conditions formation of such a state was the most probable intermediate step in double ionization \cite{Camus12prl}. Such a Coulomb complex can be  ionized subsequently later during the laser pulse \cite{sacha_pathways_2001,Camus12prl,Ho2005prl,Liu2008prl,Liu2010prl}.

A direct proof of the intermediate rescattering step in double ionization is the strong dependence of the double ionization probability on polarization ellipticity of laser pulses
(Figure 1b). Already a small ellipticity (Figure 1b, red and green curves)
steers the trajectories of the electrons away
from the parent ion and hence completely
suppresses recollision \cite{Corkum93prl,Dietrich94pra}. At higher intensities ($\unit[3\times10^{14}]{Wcm^{-2}}$, Figure 1, in blue) the double ionization probability is almost independent of the polarization ellipticity indicating sequential liberation of electrons.

\begin {figure}[t]
  \begin{center}
    \includegraphics[width=0.5\linewidth]{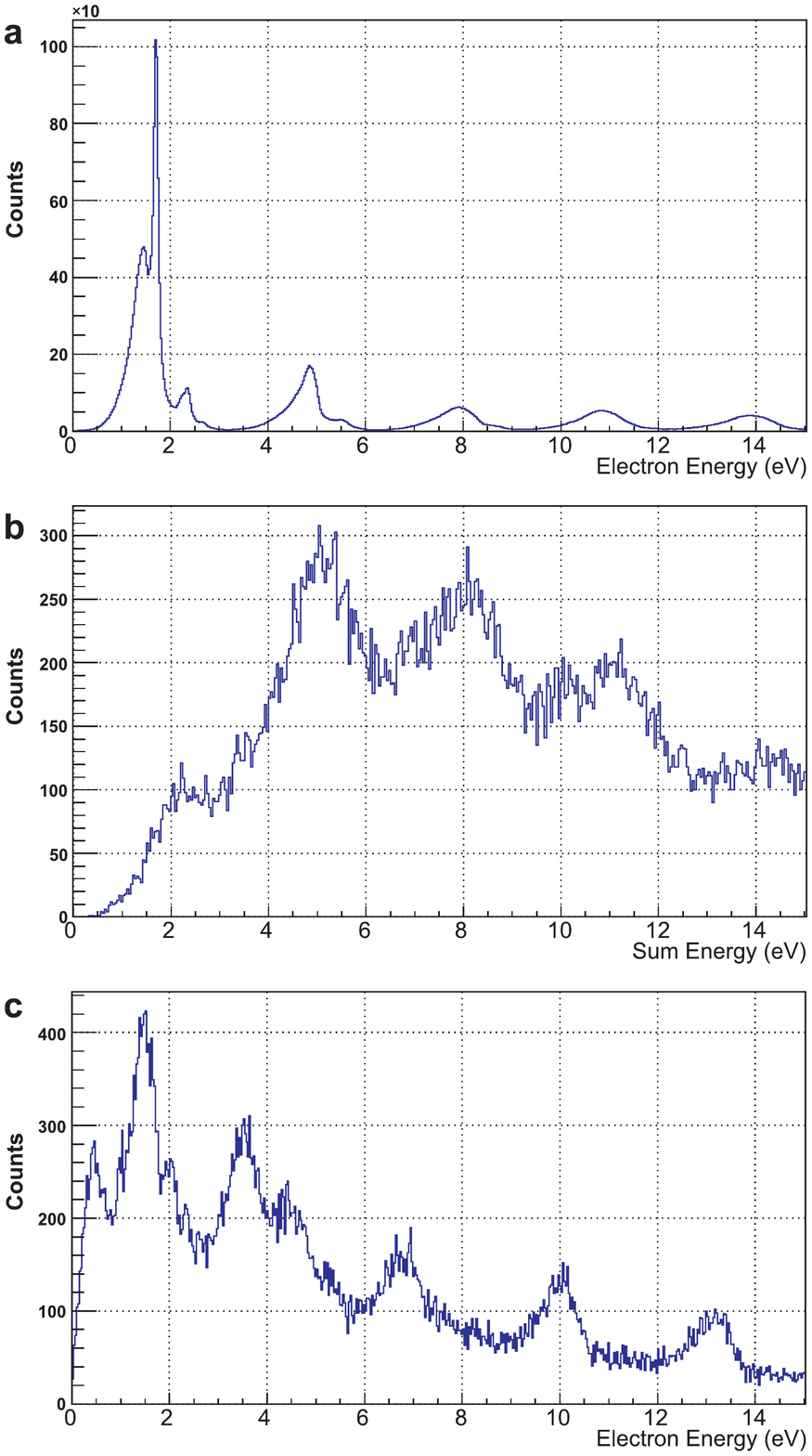}
    \caption{(a) Electron energy distribution for single ionization of Ar at \unit[394]{nm}
     $1.3\times10^{14}Wcm^{-2}$ (marked with the red arrow in Figure 1). (b) Sum of the energy of both electrons
     from double ionization of Ar at the same conditions as (a).
     (c) Energy of one of the two electrons from double ionization.
    }
  \label{fig2}
  \end{center}
\end {figure}

Having established that recollision is the
decisive step in double ionization also at \unit[394]{nm}
we now turn to the predicted ATI structure.
Figure 2 shows the electron energy spectrum for
single ionization in comparison to our findings
for double ionization. The single ionization
spectrum is in agreement with previous work
\cite{Maharjan06jpb}. It shows discrete peaks
spaced by the photon energy (\unit[3.14]{eV}). In Ar the first peak
shows a fine structure resulting from Freeman
resonances \cite{Maharjan06jpb,Freeman91jpb}.
Here the electron is first excited by resonant
multi-photon absorption to an excited field
dressed Rydberg state. As the pulse rises and
falls the intensity dependent Stark shift moves
the Rydberg states at some point in resonance
with a multiple of the photon energy. From there
the electron is then ejected by a single photon
absorption. The Stark shift of these states is
almost identical to the one of the continuum,
which leads to intensity independent peak
positions in the ATI electron spectrum. Figure 2b
displays the sum energy of both electrons
detected in coincidence with a doubly charged
Argon ion. The spectrum shows the first experimental
observation of a progression of ATI peaks in
double ionization. More surprising is however
the energy distribution of one of the two
electrons shown in Figure 2c. It shows a
pronounced multi-peak structure. The discretization of the single electron energy is expected for the case of sequential double ionization. However, under present experimental conditions sequential nature of double ionization can be excluded as was discussed above (Figure 1).

\begin {figure}[t]
  \begin{center}
    \includegraphics[width=0.5\linewidth]{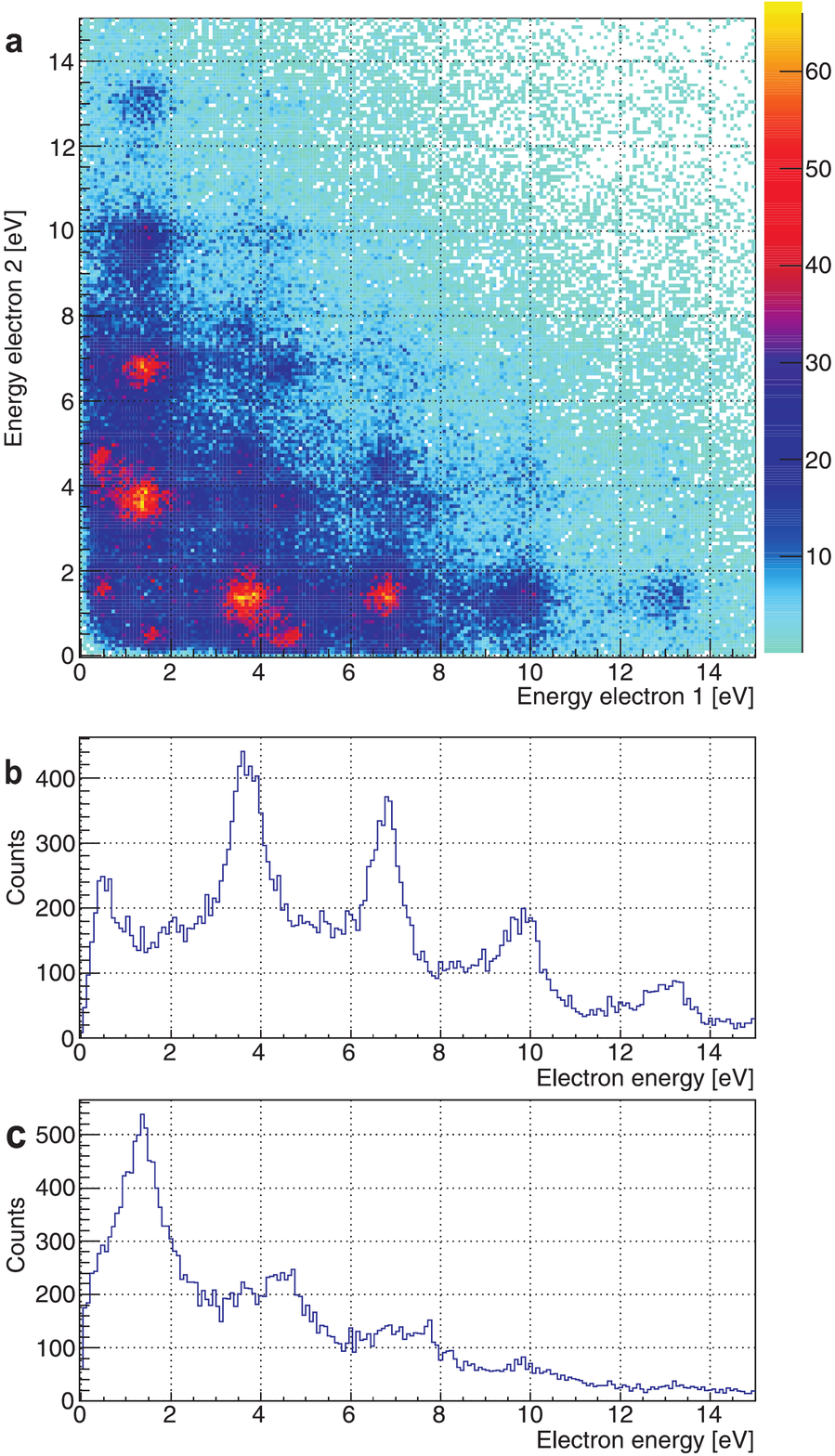}
    \caption{Double ionization of Ar by \unit[394]{nm}, \unit[45]{fsec} pulses. Same laser conditions as in Figure 2.
    (a) Energy of one electron versus the energy of the other electron. The electrons are integrated over
    all angles. (b) energy of one of the electrons if the other electron has an energy between   		\unit[1.0 - 1.6]{eV} i.e.
    projection of (a) in the range \unit[1.0 - 1.6]{eV}. (c) same as (b) for the range
    \unit[3.0 - 3.8]{eV}.
    }
  \label{fig3}
  \end{center}
\end {figure}

To shed more light onto these discrete
structures we plot the energy correlation between
the two electrons in Figure 3a. Constant sum
energy $E_1+E_2$ is found along diagonal lines in
this plot. Any recollision scenario in which the
system as a whole absorbs a discrete number of
photons (eqn 1) whose energy is then continuously
shared among the electrons would lead to counts continuously
distributed along these lines. In our data, however, the
observed electron pairs rather form islands which are
clustered along these diagonals of constant sum
energy. Thus, in addition to quantization of the
sum energy, sharing of the energy between
the electrons is not continuous either.

We now discuss the possible mechanism of the observed double ionization. The first step, as was already discussed above,
is formation of a Coulomb complex upon recollision of the first electron. Information about its preparation is believed to be lost due to a very strong electron-electron interaction upon recollision \cite{sacha_pathways_2001,Camus12prl}. The second step is liberation of both electrons from this excited compound state.
The discrete islands in Figure 3a rule out all scenarios where the
electron pair is set free into the continuum
simultaneously, i.e. during the same laser half-cycle. In any such scenario the electrons would
originate spatially close to each other. In this
case the electron-electron repulsion would add a
continuous kinetic energy to both electrons. For
instance, the potential energy stored in the
electron-electron repulsion at a distance of \unit[5]{a.u.} would already translate to an energy of \unit[2.7]{eV} on each of the electrons, sufficient to wash
out any narrow peaks. It is known from single photon double ionization that the nearly simultaneous birth of the electron pair leads to a smooth energy spectrum \cite{Wehlitz91prl}.

If, however, the two electrons were ejected
sequentially from a Coulomb complex, then the two emission steps
would each produce their own ATI progression. Each
such ATI spectrum would have the same spacing
between the peaks, but could have different
offsets. Close inspection of the pattern in
Figure 3a shows indeed, that it consists of
mainly two ATI spectra. We obtain these two ATI
spectra by projecting horizontal slices from the
data shown in Figure 3a. This procedure yields
electron spectra of electron 1 for fixed energy of
electron 2. These two ATI spectra show a
different offset and a different envelope. All
islands in Figure 3a are located at points
corresponding to one electron being from the ATI
sequence in Figure 3b and the other electron from
the shifted one in Figure 3c. Therefore, one of the plausible explanations of the present results would be an out-of-phase scenario, as predicted by the classical simulation \cite{Ho2005prl}, where the electrons are liberated at different phases of the laser field after multiple recollisions. In this case the electrons are emitted in opposite directions, as was experimentally observed on argon under similar laser field conditions ($3.17\times{U_p}\approx\unit[7.4]{eV}$), however at a wavelength of \unit[800]{nm} \cite{Liu2008prl}. 

Presently light driven double ionization is
discussed heavily in three rather different
regimes of complexity. The single photon case,
which is mostly well understood in atoms \cite{Ni11jpb} and
molecules (see e.g. \cite{Akoury2007sci}), the two photon case as a central topic at
Free Electron Laser science (see e.g. \cite{Rudenko2010jpb}) and finally the
strong field case discussed here. The observed
energy discretization shows a direct link
between these fields. A discrete energy of the
photon field is one of the quantum effects common
to all three problems. It is related to a second
quantum effect, which is the symmetry of the many
particle wavefunction in the continuum. This is governed by the
Pauli principle. To access the role of symmetry
the photon number and hence the total parity of
the final state has to be known \cite{Ni11jpb}.
For the one and two photon case this symmetry is
the most important ingredient shaping the two
electron continuum (see
\cite{briggs00jpb,Ni11jpb}). The ability to count
the photons even for a multi-photon multi-electron process paves the road to explore these
symmetries in the future.

\textbf{Acknowledgment} The experimental work was
supported by the Deutsche Forschungsgemeinschaft.
We thank Jonathan Parker, Andreas Becker, Manfred
Lein, Klaus Bartschat und Oleg Zatsarinny for
helpful discussion.

\bibliographystyle{unsrt}

\bibliography{bibliography}

\end{document}